\documentstyle[12pt]{article}
\newcommand{\be}{\begin{equation}}
\newcommand{\ee}{\end{equation}}
\newcommand{\bea}{\begin{eqnarray}}
\newcommand{\eea}{\end{eqnarray}}
\newcommand{\beq}{\begin{equation}}
\newcommand{\eeq}{\end{equation}}
\newcommand{\beqa}{\begin{eqnarray}}
\newcommand{\eeqa}{\end{eqnarray}}
\newcommand{\beqar}{\begin{eqnarray*}}
\newcommand{\eeqar}{\end{eqnarray*}}
\newcommand{\labell}[1]{\label{#1}\qquad_{#1}} 
\def\l{\lambdabel}
\def\o{\over}
\def\a{\alpha}
\newcommand{\beas}{\begin{eqnarray*}}
\newcommand{\eeas}{\end{eqnarray*}}

\begin{document}

\makeatletter
\renewcommand{\theequation}{\thesection.\arabic{equation}}
\@addtoreset{equation}{section}
\makeatother

\begin{titlepage}

\vfill

\begin{center}
   \baselineskip=16pt
   {\Large\bf  On EVH black hole solution in Heterotic string theory}
   \vskip 2cm
    Hossein Yavartanoo
       \vskip .6cm
             \begin{small}
      \textit{Department of Physics, Kyung Hee University, \\ Seoul 130-701, Republic of Korea}
        \end{small}\\*[.6cm]
\end{center}

\vfill
\begin{center}
\textbf{Abstract}\end{center}
We study the near horizon geometry of charged rotating black holes in toroidal compactifications of heterotic string theory.  We analyze  the extremal vanishing horizon (EVH) limit for these black hole solutions and we will show that the near horizon geometry develops an AdS$_3$ throat. Furthermore, we will show that the near horizon limit of near EVH black holes has a  BTZ factor.  We also comment on the CFT dual to this near horizon geometry. 
\begin{quote}
\end{quote}
\vfill
\end{titlepage}
\section{Introduction}
Black holes are among the most intriguing and fascinating objects in the Physics landscape. Every black hole has an entropy, associated with the area of its horizon.  This implies that a black hole is hot and can radiate with the frequency spectrum which is characteristic of a blackbody. This classical thermodynamics consideration suggests the existence of quantized microstates, whose degeneracy would account for the macroscopic production of the black hole entropy.  The quantum theory of gravity would be expected to explain the nature of these microstates, thereby solving the black hole entropy problem.   

Strominger and Vafa in a celebrated paper \cite{Strominger:1996sh} have shown that string theory techniques can be used to count the quantum microstates associated to classical black hole configurations. This result has been generalized to many different black hole solutions in different dimensions as well as to ones that are near extremal. 

Although a complete counting of the quantum microstates of a generic black hole is not achieved yet, one for which we presumably need a complete understanding of quantum theory of gravity, such a counting has been obtained for many black hole solutions by identifying their underlying microstates with those of a dual two or higher dimensional conformal field theory (CFT). In most of the examples in which the identification has been done, black holes possess an AdS$_3$ throat in their near horizon limit and the degeneracy of their microstates can be captured by a two dimensional CFT using AdS$_3$/CFT$_2$ duality.

Besides black holes with AdS$_3$ throats, there are some recent proposals towards the identification of microstates of extremal black holes. The near horizon geometry of general extremal (but not necessarily supersymmetric) black holes contains an AdS$_2$ throat \cite{Kunduri:2008rs}. This fact, if we have a formulation of AdS$_2$/CFT$_1$ duality (see \cite{Sen-AdS2/CFT1} for a review on progress in this direction), may be used for giving a statistical account of general extremal black hole entropy.

It has been conjectured   \cite{Kerr-CFT}  that an extremal Kerr black hole is dual to a chiral two dimensional CFT.  The conjecture has been extended to many other extremal black hole solutions  \cite{Ext/CFT} (for a recent review of related discussions see  \cite{Simon:2011zza} ). Although the Kerr/CFT conjecture is very interesting, to be precise, it is rather a suggestion for a possible pair of theories dual to each other and many things should be understood to establish the proposal as a concrete duality.  A precise identification of the proposed chiral CFT is still an open question and there have been arguments that the Extremal/CFT proposal does not have the same dynamical content as the standard AdS/CFT and may only be used for reading the entropy (see   \cite{Kerr-CFT-caveats} for a discussion on this point).  It has been discussed that these AdS$_2$ geometries do not generically represent a decoupled conformal field theory. Even if they do, the AdS/CFT machinery would suggest these theories may be dynamically trivial  \cite{Sen-AdS2/CFT1,DLCQ-CFT}, in the sense that they only contain degeneracy of the vacuum in their spectrum.

Using the Kerr/CFT approach,  a large class of  rotating black hole solutions in four dimensions have been studied in \cite{Chow:2008dp} (see also \cite{Shao:2010cf, Mei:2010wm, Ghezelbash:2009gf} for related discussions).  In particular a black hole which carries four charges and known as the Cvetic-Youm solution has been studied in some details.   It has been shown that there exists a hidden conformal symmetry corresponding to the near horizon geometry and that the Cardy formula for these near-horizon geometries is satisfied as well.  The microscopic entropies of the dual CFTs agree with the BekensteinÐHawking entropies of the extremal rotating black hole.

In a more recent approach, a specific class of extremal black holes, the {\it Extremal Vanishing Horizon} (EVH) black holes is studied in \cite{SheikhJabbaria:2011gc}. In particular the general 4d EVH black
hole solutions have been studied in Einstein gravity coupled to a scalar and a gauge field. For a general stationary black hole solution in this theory it has been shown that the near horizon limit of any EVH black hole, if it exists, has  an AdS$_3$ throat. Using this observation the EVH/CFT correspondence is proposed: gravity on the near horizon of the EVH geometry is described by a 2d CFT ( for some recent discussions see   \cite{deBoer:2011zt, Yavar}). 

In this paper, we use the approach given in \cite{SheikhJabbaria:2011gc} for studying the near horizon geometry of the four dimensional Cvetic-Youm black holes. In the next section we review the  Cvetic-Youm black hole solution and its thermodynamics.  In Section 3 we analyze the EVH limit for this black hole solution. In Section 4 we will study the near horizon geometry of the EVH black hole and we will show that the near horizon geometry develops an AdS$_3$ throat. This AdS$_3$ is however, generically a pinching AdS$_3$. Furthermore, we will show that near horizon limit of near EVH black holes has a pinching BTZ factor.  The appearance of the AdS$_3$ factor in the near horizon geometry is a good omen for trying to establish the EVH/CFT for this black hole solution. We will comment on the CFT dual to this near horizon geometry in Section 5. Section 6 is devoted to the conclusion. 
\section{Rotating Black Hole Solution in Heterotic String Theory}
We start with toroidal compactification of heterotic string theory to four dimensions. At the low energy limit,
the theory consists of 
gravity coupled to a complex scalar $S=S_1+iS_2$,
a 4$\times 4$
matrix valued scalar field $M$ satisfying the constraint

\be
MLM^T = L, \qquad L=\pmatrix{0 & { 1}_{2\times 2} \cr { 1}_{2\times 2}  & 0}\,,
\ee
along with four U(1) gauge fields $A_\mu^{(i)}$
($i=1\cdots 4$).\footnote{Actual heterotic string theory has 28 gauge
fields and a 28$\times$28 matrix valued scalar field, but the
truncated theory discussed here contains all the non-trivial information
about the theory.}  The
bosonic part of the Lagrangian density is
\bea
\label{Lagrangian}
{\mathcal L} &=& R - {1\over 2}
\, g^{\mu\nu} S_2^{-2} \partial_{\mu} \bar S \partial_{\nu} S  + {1\over 8} g^{\mu\nu}
Tr(\partial_{\mu} M L \partial_{\nu} M L) \nonumber \\
&& -
{1\over 4}\, S_2 g^{\mu\rho}
g^{\nu\sigma}
F^{(i)}_{\mu\nu} (LML)_{ij} F^{(j)}_{\rho\sigma}
+ {1\over 4}\, S_1 g^{\mu\rho}
g^{\nu\sigma}
F^{(i)}_{\mu\nu} L_{ij}
\widetilde F^{(j)}_{ \rho\sigma}\, ,
\eea
where
\be
\widetilde F^{(i)\mu\nu} = {1\over 2}\, (\sqrt{-\det g})^{-1}
\epsilon^{\mu\nu\rho\sigma} \, F^{(i)}_{ \rho\sigma}\, .
\ee
General rotating black hole solutions in this theory, with electric charge
vector $\vec Q$ and magnetic charge vector $\vec P$, have been
constructed in \cite{Cvetic:1996kv}. 
\be
Q = \pmatrix{ 0 \cr Q_2 \cr 0 \cr Q_4}\, ,
\qquad P=\pmatrix{P_1 \cr 0 \cr P_3\cr
0}
\, .
\ee
These black hole solutions break all the supersymmetries of the theory.
It is more convenient to parametrize the matrix valued scalar
field $M$ as
\be
M =\pmatrix{G^{-1} & -G^{-1} B\cr B G^{-1} & G - B G^{-1} B}
\ee
where $G$ and $B$ are $2\times 2$ matrices of the form
\be
G=\pmatrix{G_{11} & G_{12} \cr G_{12} & G_{22}}, \qquad
B=\pmatrix{0 & B_{12} \cr -B_{12} & 0}\, ,
\ee
where $G$ and $B$ represent components of the string metric and
the anti-symmetric tensor field along an internal two dimensional
torus.
The solution is given by
\bea
\label{solution}
G_{11}&=&{{(r+2s_4^2)(r+2s_2^2)+m^2j^2{\rm cos}^2 \theta}\over {(r+2s_3^2)(r+2s_2^2)+m^2j^2{\rm cos}^2\theta}},  \cr
G_{12}&=&{2j{\rm cos}\theta(s_3c_4s_1c_2-c_3s_4c_1s_2)\over{(r+2s_3^2)(r+2s_2^2)+m^2j^2{\rm cos}^2\theta}},  \cr
G_{22}&=&{{(r+2s_3^2)(r+2s_1^2)+m^2j^2{\rm cos}^2 \theta}\over {(r+2s_3^2)(r+2s_2^2)+m^2j^2{\rm cos}^2\theta}},  \cr
B_{12}&=&-{{2j{\rm cos}\theta(s_3c_4c_1s_2-c_3s_4s_1c_2)}\over{(r+2s_3^2)(r+2s_2^2)+m^2j^2{\rm cos}^2\theta}},  \cr
Im\, S&=&{\Delta^{1\over 2}\over{(r+2s_3^2)(r+2s_4^2)+m^2j^2{\rm cos}^2 \theta}},\cr
ds^2&=&\Delta^{1\over 2}\bigg[-{{r^2-2mr+m^2j^2{\rm cos}^2\theta}\over\Delta}dt^2+{{dr^2}\over{r^2-2mr+m^2j^2}} + d\theta^2 \cr &&    
+{{{\rm sin}^2\theta}\over \Delta}\{\prod_{i=1}^4(r+2s_i^2)+m^2j^2(1+{\rm cos}^2\theta)r^2+W  + 2m^3j^2r{\rm sin}^2\theta\}d\phi^2 \cr &-&
{{4j\sin^2\theta}\over \Delta}\left((\prod_{i=1}^4 c_i - \prod_{i=1}^4 s_i)r+2m \prod_{i=1}^4 s_i\right)\theta dtd\phi\bigg], 
 \eea
where
\bea
W &\equiv& 2m^2j^2 r  \sum_{i=1}^4 s_i^2  + 8j^2\left( \prod_{i=1}^4 s_ic_i -\prod_{i=1}^4 s_i^2 - m \sum_{k=1}^4 \frac{\prod_i s_i^2}{2s_k^2}\right)
+m^4j^4{\rm cos}^2 \theta, \cr
\Delta &\equiv& \prod_{i=1}^4(r+2s_i^2) + (2m^2j^2r^2+W){\rm cos}^2\theta, \hspace{3mm} s_i\equiv \sqrt{m}\sinh\delta_i, \hspace{3mm} c_i\equiv \sqrt{m}\cosh\delta_i\nonumber
\eea
The axion field $\Re S$, and gauge fields  also vary with spatial coordinates, but these expressions turn out to be cumbersome and we do not present them here explicitly. 

The solution is parametrized by six parameters $j$, $m$, $\delta_{1}$, $\delta_{2}$, $\delta_{3}$ and $\delta_4$. For later purpose, let us rewrite the above metric into the following form
\be 
ds^2=-N^2dt^2 + g_{rr}dr^2 + g_{\phi\phi}\left(d\phi+N^{\phi}dt\right)^2+g_{\theta\theta}d\theta^2
\ee 
where $N^2=-g_{tt}+g_{\phi\phi}\left(N^{\phi}\right)^2$     and $g_{t\phi}=g_{\phi\phi}N^{\phi}$.  By eliminating the conical singularity in the Euclidean $(\tau= it,r)$ sector, we obtain the black hole temperature
\be
\label{BHtemp}
T=\frac{1}{\Delta \tau} = \left(\frac{\left(N^2\right)'}{4\pi\sqrt{g_{rr}N^2}}\right)_{r=r_+}
\ee
where {\it prime} means derivative with respect to $r$ and $r_+$ is the location of the outer horizon. Denoting by $r_+$ and $r_-$ the location of inner and outer horizons we can factorize $N$, $N^{\phi}$ and $g_{rr}$ as follows
\bea
&& N^2=(r-r_-)(r-r_+)\mu(r,\theta) \nonumber \\
&& N^{\phi}= -\omega+(r-r_+)\eta(r,\theta) \nonumber \\
&& g_{rr}=\frac{1}{(r-r_-)(r-r_+)\Lambda(r,\theta)} \nonumber
\eea 
where $\omega$ is the angular velocity of the horizon and we assume functions $\mu$, $\Lambda$ and $\Lambda$ do not have zero in $(r_+,\infty)$. The inner and outer horizon are given by
\be \label{horizons}
r_{\pm}=m(1\pm \sqrt{1-j^2})
\ee 
 For solution (\ref{solution}) the black hole temperature which reads off from (\ref{BHtemp}) is given by
\be 
\label{temp}
T=\frac{r_+^2-r_-^2}{16\pi \left( r_+\prod_i c_i+ r_- \prod_i s_i \right)}
\ee

The ADM mass $M$,  electric and magnetic charges $\{Q_i,P_i\}$,
and the angular momentum $J$ are given by:
\bea
&& Q_2 =4 c_1s_1, \hspace{3mm} Q_4 = 4 
c_2s_2, \hspace{3mm} 
P_1 = 4
c_3s_3, \hspace{3mm} P_3 = 4
c_4s_4,\cr
&& G_4 M={1\over 2}  \sum_{i=1}^4 c_i^2-m, \hspace{9mm} G_4 J=   j( \prod_{i=1}^4 c_i -\prod_{i=1}^4 s_i).
\eea
The Bekenstein-Hawking entropy  associated with this solution turns out to be
\be 
\label{entropy}
S_{BH} = \frac{A_h}{4G_4} = \frac{4
\pi\left( r_+ \prod_{i} c_i  + r_- \prod_{i} s_i\right) }{G_4(r_++r_-)}
\ee

The extremal limit where black hole temperature vanishes is given by limit $r_+\rightarrow r_-$, and from  (\ref{horizons}) it is evident that we have two different kinds of extremal limit corresponding to the limits $j=1$ and $m=0$ and we shall denote these by ergo-branch and ergo-free branch respectively.\footnote{In the extremal limit in the ergo-free branch we should take one
or three of the $\delta_i$'s negative, and then taking the limit $|\delta_i|\to
\infty$, $m\to 0$, in a way that keeps the $Q_i$, $P_i$ and
$J$ finite. }
\section{EVH limit of the Black hole solution}
The generic  black hole solution (\ref{solution}) is a stationary geometry. In the extremal case, the Kerr/CFT-type analysis has been carried out and shown that in these cases there are different U(1) isometries which enhance to chiral Virasoro of the proposed dual chiral 2d CFT \cite{Ghezelbash:2009gf}.  In particular it has 
been argued that by choosing the proper boundary conditions for the gravitational field, dilaton and gauge fields one can find the diffeomorphisms that generate Virasoro algebra without any central charge. The generator of diffeomorphisms which is a conserved charge, can be used to construct an algebra under Dirac brackets. This algebra is the same as the diffeomorphism algebra but just with some extra central terms. These central terms, in general contribute to the central charge of the Virasoro algebra, but it has been shown that the only non-zero contribution to the central charge of the dual conformal field theory comes from the gravitational field.  The central charge together with Frolov-Thorne temperature give us  the microscopic entropy of the extremal rotating  black hole in dual chiral CFT. The microscopic entropy is exactly the same as macroscopic Bekenstein-Hawking entropy of the extremal black hole (\ref{temp}).

In \cite{SheikhJabbaria:2011gc} it has been shown that one can get AdS$_3$ throat as the near horizon of geometry of a rotating black hole in a special subspace of the black hole moduli space.  This is often corresponding to the large limit of charges carried by the black hole. This issue is discussed with some details in a recent paper \cite{deBoer:2011zt}.  For a given temperature this large charge limit turns out to be an infinite entropy. To keep the black hole entropy finite, we should combine this limit with vanishing horizon limit. This implies temperature vanishes at this limit too.  

In this section we analyze the near horizon geometry of  EVH black hole in the low energy limit of heterotic string theory which is described by the Lagrangian (\ref{Lagrangian}). The general analysis for a rotating black hole in Einstein-Maxwell-dilaton theory has been done in \cite{SheikhJabbaria:2011gc}.  Here we follow the same line to get the EVH solution.  Therefore we are seeking the limit on the moduli space of the black hole, which in this case is parametrized by six parameters $m, j$ and $\delta_i$,  in which the area of the horizon and temperature vanishes  while the ratio remains finite. 
The EVH black hole parameter space is then a four dimensional hypersurface in this parameter space, associated with $A_h = 0, T = 0$. This hypersurface is called the EVH hypersurface. Each point on the EVH hypersurface corresponds to an EVH black hole. We will also consider moving slightly away from the EVH hypersurface, then by definition, we get a near EVH black hole with small $A_h$ and $T$ but with fixed $A_h/T$. The near EVH black hole is hence specified by two parameters around a given EVH point.

To study the EVH limit of a given black hole, we notice that for any black hole solution, entropy is a positive-definite function of charges and temperature and the entropy can vanish only at zero temperature. Therefore, we consider the low temperature  expansion for the Bekenstein-Hawking entropy of a black hole as follows
\be
S(q_i,T) = S_0(q_i) + S_1(q_i)T + S_2(q_i)T^2 + \cdots
\ee
where $q_i$ stand for the different black hole charges. Generic extremal black holes solutions have non-zero $S_0(q_i)$, providing the dominant contribution to the entropy in the near extremal limit. For the black hole solution (\ref{solution}), the low temperature expansion of the Bekenstein-Hawking entropy (\ref{entropy}) is given by
\be
\label{enex}
S_{\mathrm{BH}}= \frac{2\pi}{G_{\mathrm{N}}}\left(  \prod_{i} c_i  + \prod_{i} s_i \right) + \frac{8\pi^2}{G_{\mathrm{N}} r_+}  \left( \prod_{i} c_i^2  - \prod_{i} s_i^2\right)  T +\cdots
\ee

We are seeking a family of extremal black holes for which the coefficient $S_0$ is zero. In that situation, the leading contribution to the entropy is $S \sim S(q_i)T$ and one may speculate on the existence of a dual two dimensional CFT, since $S  \simeq cT^k$ follows from conformal invariance with $c$ being some effective central charge.  A similar low temperature expansion and similar reasoning also applies to the near-BPS black p-brane solutions. In general one gets $S\sim c T^k$ in the low energy expansion of the  Bekenstein-Hawking entropy ,which for $k = 2,3,5$ leads to the usual (maximally supersymmetric) AdS$_{k+2}$/CFT$_{k+1}$ examples.

From (\ref{enex}), it is clear that $S_0$ is non-vanishing unless one or three of the $\delta_i$Õs are negative.  In addition to get the second term non-zero we need to take the limit $r_+\rightarrow 0$ with the same rate as $S_0$ vanishes. This needs to scale $j\rightarrow 1$ and $m\rightarrow 0$ while keeping $m/\sqrt{1-j^2}$ finite. In addition to keep the solution non-trivial we need to take $c_i, s_i$'s finite, which implies taking $\delta_i\rightarrow \infty$.  This procedure can  be summarized as follows
\be 
\label{EVHlimit}
m=\mu\epsilon, \quad  1-j^2=l^2\epsilon^2,  \quad \epsilon\rightarrow 0,
\ee  
with keeping $s_i,  l$ and $\mu$ finite. In this limit the black hole entropy and temperature scale to zero with the same rate as $\epsilon$.  

\be
\label{ST}
S_{\mathrm BH}= \frac{\pi(2\mu \lambda +l R_{{\mathrm AdS}_3}^4) }{4R_{{\mathrm AdS}_3}^2 G_4}\;\epsilon ,\quad\quad T=\frac{2\mu l R_{{\mathrm AdS}_3}^2}{\pi(2\mu \lambda +l R_{{\mathrm AdS}_3}^4)}\epsilon
\ee
where 
\be 
\lambda=32\sum_{i<j<k}s_i^2s_j^2s_k^2, \qquad  R_{{\mathrm AdS}_3}^2= 16|s_1s_2s_3s_4|.
\ee 

In other words, in the six dimensional parameter space of rotating black hole solutions, there exists a four dimensional EVH hypersurface parameterized by $s_i$.  The EVH black hole metric is then obtained by setting $m = 0, j = 1$, for which the metric takes the form
\be
\label{metricEVH}
ds^2=-N^2dt^2+g_{rr}dr^2+g_{\phi\phi}\left(d\phi+N^{\phi}dt\right)^2+g_{\theta\theta}d\theta^2 
\ee
where functions $N^2, g_{rr}, g_{\phi\phi}, N^{\phi}$ and $g_{\theta\theta}$ are given by

\be
\label{EVHBH}
N^2=\frac{r h}{f}, \quad g_{rr}=\frac{h}{r^2},  \quad g_{\phi\phi}=\frac{rf \sin^2\theta}{h}, \quad g_{\theta\theta} = h, \quad N^{\phi}=\frac{4s_1s_2s_3s_4}{f},
\ee
where functions $f$ and $h$ are defined by
\be
f=r^3+2r^2\sum_i s_i^2 + 4r \sum_{i<j}s_i^2s_j^2 +8 \sum_{i<j<k}s_i^2s_j^2s_k^2,\quad h=\left( r f + 16 s_1^2s_2^2s_3^2s_4^2 \sin^2\theta  \right)^{1/2} .
\ee
The horizon of the above EVH  black hole is located at $r = 0$. The location of singularity $r_s$ however is given by the zeros of $h$:
\be\label{sing}
r_s = r_s(\theta), \quad \prod_i(r_s+2s_i)-16\cos^2\theta\prod_i s_i^2 = 0.
\ee
From (\ref{sing}) we observe that for generic values of $\theta$ the singular line lies in the $r < 0$ region. More precisely, $r_s \leq 0$ and equality happens for $\theta = 0, \pi$ , that is $r_s(0) = r_s(\pi) = 0$. This is the picture which is discussed in \cite{SheikhJabbaria:2011gc}. The singularity which is located at negative $r$ is generically sitting behind the horizon which is at $r = 0$. The singularity becomes ÒnakedÓ only in two points: $r = 0,\theta = 0$ and $r = 0,\theta = \pi$. For the EVH black hole (\ref{EVHBH}), therefore, away from these two singular points, the horizon is generically far from the singularity, and indeed we define our near horizon limit such that, generically, we are parametrically infinitely far from the singularity. This is how a general EVH  black hole is different from the Òsmall black holesÓ where the singularity and horizon are always arbitrarily close.  

For small black holes in string theory it has been shown that adding the higher derivative corrections blows up the horizon to non-zero size and the resulting Bekenstein-Hawking entropy precisely matches with counting the corresponding microstates. It is interesting to study the effect of higher derivative corrections to the horizon shape for EVH black holes\footnote{For a related discussion see \cite{myworks}}.

Moreover, from the above metric one can find the geometric shape of the horizon of the EVH black hole. This is topologically a two-sphere, but a singular one, because $g_{\phi\phi}$ vanishes at the horizon $r=0$. In other words, close to the horizon and at constant $r$ and $t$ the metric is more like a cylinder, the axis of which is along $\theta$ direction and its circle, which has vanishing radius is along $\phi$ direction.
\section{Near horizon limit of EVH black hole}
One may study the near horizon limit of the geometry obtained in the EVH limit. In order to do that, let us consider limit (\ref{EVHlimit}) and apply the scaling
\be 
\label{scales}
r=(\mu+\frac{R_{{\mathrm AdS}_3}^4}{\lambda}\rho^2)\epsilon,\quad t=\frac{\lambda}{R_{{\mathrm AdS}_3}^2}\frac{\tau}{\sqrt{\epsilon}},\quad  \phi= \frac{\psi+\tau}{\sqrt{\epsilon}}
\ee 
with $\rho$, $\tau$ and $\psi$ held fixed. In this limit, the metric (\ref{metricEVH}) takes the form
\be
\label{metricNHEVH}
ds^2=R_{{\mathrm AdS}_3}^2\sin\theta\left[ -\rho^2d\tau^2 +\frac{d\rho^2}{\rho^2}+\rho^2d\psi^2+\frac{1}{4}d\theta^2      \right].
\ee

All gauge field strengths vanish and the scalar fields are given by
\be
\label{scalars1}
G_{11}=\frac{s_4^2}{s_3^2},\quad G_{22}=\frac{s_1^2}{s_2^2}, \quad G_{12}=-\left|\frac{s_1s_4}{s_2s_3}\right|\cos\theta,\quad  B_{12}=-\left|\frac{s_1s_4}{s_2s_3}\right|\cos\theta,
\ee 
\be
\label{scalars2}
S_1=\left|\frac{s_1s_2}{s_3s_4}\right|\cos\theta,\quad S_1=\left|\frac{s_1s_2}{s_3s_4}\right|\sin\theta.\quad
\ee
As we see the near horizon metric (\ref{metricNHEVH}) is exactly of
the form that was outlined and discussed in \cite{SheikhJabbaria:2011gc}. In this case, however, the AdS$_3$ radius $R_{{\mathrm AdS_3}3}$ and the value of the dilaton fields are determined by the value of the charges $Q_i, P_i$, defining the EVH black hole. Although not implied by the equations of motion on the near horizon geometry, the value of all parameters of the near horizon configuration are fixed by the charges defining the full EVH black hole, once it is extended out of the horizon and to the asymptotic flat region. In this sense, the EVH black hole shows Òattractor behaviourÓ.

One may also study the near horizon limit of near EVH black hole. To this end, let us consider  limit (\ref{EVHlimit})  together with the following scaling
\bea
r=\mu\epsilon+\frac{R_{{\mathrm AdS}_3}^4}{\lambda}\left(\rho^2-\frac{1}{4}l^2-\frac{\mu^2\lambda^2}{R_{{\mathrm AdS}_3}^8} \right)\epsilon^2, \quad t=\frac{\lambda}{R_{{\mathrm AdS}_3}^2}\frac{\tau}{\epsilon}, \quad \phi=\frac{1}{\epsilon }(\tau+\psi),
\eea
Taking the limit $\epsilon \rightarrow 0$, we obtain the following geometry
\be
ds^2=R_{{\mathrm AdS}_3}^2|\sin\theta|\left(-F(\rho)d\tau^2+\frac{d\rho^2}{F(\rho)}+\rho^2(d\psi-\frac{\rho_+\rho_-}{\rho^2}d\tau)^2+\frac{1}{4}d\theta^2\right)
\ee
where \be F(\rho)=\frac{(\rho^2-\rho_+^2)(\rho^2-\rho_-^2)}{\rho^2}, \ee 
and  $\rho_{\pm}$ are given by
\be
\rho_+=\frac{1}{2}\left| l+\frac{2\lambda \mu}{R_{{\mathrm AdS}_3}^4}\right|,\quad \rho_-=\frac{1}{2}\left| l-\frac{2\lambda \mu}{R_{{\mathrm AdS}_3}^4}\right|
\ee
The gauge fields vanish and scalar fields take the same values as in (\ref{scalars1}) and(\ref{scalars2}). The pinching AdS$_3$ in (\ref{metricNHEVH}) is now replaced by a pinching BTZ. The Bekenstein-Hawking entropy of the pinching BTZ solution to this 3d theory is then\footnote{One may reduce the 4d theory (\ref{Lagrangian}) over the metric ansatz
$ds^2 = R_{{\mathrm AdS}_3}^2 | \sin\theta | \left(g_{ab}dx^adx^b +\frac{1}{4}d\theta^2\right)$ with $a, b = 1, 2, 3$ and $\theta \in [0, \pi]$. In the gravity sector of the 3d reduced action we obtain an AdS$_3$ theory with 3d cosmological constant $R_{{\mathrm AdS}_3}^{-2}$ and 3d Newton constant $G3 = \frac{G4}{R_{{\mathrm AdS}_3}}$ .}
\be
S_{\mathrm 3d}=\frac{2\pi\epsilon \,\rho_+}{4G_3} =\frac{\pi(2\mu\lambda + lR_{{\mathrm AdS}_3}^4 )\epsilon}{4G_3R_{{\mathrm AdS}_3}^4} = \frac{\pi(2\mu \lambda +l R_{{\mathrm AdS}_3}^4)\epsilon }{4G_4 R_{{\mathrm AdS}_3}^2 }
\ee
which is entropy of the original black hole (\ref{ST}). It is also useful to compare the Hawking temperatures of the original near EVH black hole and that of the pinching BTZ:
\be 
T_{\mathrm BTZ}=\frac{\rho_+^2-\rho_-^2}{2\pi\rho_+}=\frac{2\lambda \mu l}{\pi(2\mu\lambda + lR_{{\mathrm AdS}_3}^4 )} =\frac{\lambda}{R_{{\mathrm AdS}_3}^2 \epsilon} \;T_{4d}
\ee 
Up to a prefactor,  this is the same as  the Hawking temperatures of the original black hole  (\ref{ST}). Recall from \ref{scales}) that this prefactor  is expected.
\section{EVH/CFT Correspondence} 

In the previous section we have studied near horizon limit of (near) EVH black holes and shown that we generically obtain an AdS$_3$ throat.  We also have seen that the entropy of the original near EVH black hole is parametrically equal to the entropy of the BTZ geometry obtained in the near horizon limit. 
The appearance of the AdS$_3$ throat in the near horizon of the EVH black hole is very suggestive of the existence of a 2d CFT dual to physics on this geometry.  However we should note that, what we obtain in the near horizon is not a round  AdS$_3$, it is a {\it pinching orbifold} of AdS$_3$. So, we first need to have proposals for ``resolving the pinching orbifold''.  In \cite{SheikhJabbaria:2011gc}, it has been argued that, the pinching can be removed by scaling Newton coupling constant $G_4\rightarrow 0$. In another word we have to accompany the already Òdouble scaling near EVH near horizon limitÓ of the previous sections by $G_4 = \epsilon b^2$, with  $b$, and  $R_{{\mathrm AdS}_3}$ = fixed. The dual 2d CFT, after resolution of the pinching orbifold singularity, has a finite central charge $c$
\be
c = \frac{3R_{{\mathrm AdS}_3}}{2b^2} .
\ee

The identification of $L_0$ and $\bar L_0$ in terms of the BTZ parameters can be done in the standard way
    \cite{DLCQ-CFT}, i.e.
\be\label{L0-barL0}%
L_0=\frac{c}{24}\left(\frac{\rho_++\rho_-}{R_{{\mathrm AdS}_3}}\right)^2\,,\qquad
\bar L_0=\frac{c}{24}\left(\frac{\rho_+-\rho_-}{R_{{\mathrm AdS}_3}}\right)^2\,,
\ee%
The BTZ black hole is then a thermal state
in the 2d CFT specified above at temperature $T_{\mathrm BTZ}= \frac{\rho_+^2-\rho_-^2}{2\pi \rho_+ }$. With this identification and recalling our earlier discussions, it is then obvious that the Cardy formula which produces the BTZ black hole entropy correctly reproduces the near EVH black hole entropy.
\section{Conclusion } 
In this paper we have analyzed the near horizon geometry of the four dimensional Cvetic-Youm black hole.  This is a charged rotating black hole solution in toroidally compactified heterotic string theory. The solution is parametrized by mass, angular momentum, two electric and two magnetic charges. Beside the four $U(1)$ gauge fields, the background includes several scalar fields which represent components of the string metric, dilaton and the anti-symmetric tensor field along an internal two dimensional torus. By analyzing the near horizon field configuration at the vanishing limit of the area of horizon and Hawking temperature we have found that geometry develops an AdS$_3$ throat. The same approach has been used in \cite{SheikhJabbaria:2011gc} to study a general rotating black hole in Einstein gravity coupled to one gauge field and one scalar field.   The AdS$_3$ throat that appears in the near horizon limit is a pinching AdS$_3$= AdS$_3$/$Z_K$ ,$ Z \rightarrow \infty$. Furthermore, we showed that near horizon limit of near EVH black holes has a pinching BTZ factor.  The appearance of an AdS$_3$ factor in the near horizon geometry is a good indication for trying to establish the EVH/CFT. It is easy to check that  our near horizon limit is indeed a decoupling limit.  This is analogous to what happens in the usual Dp-brane case in the decoupling limit \cite{{Alishahiha:2000qf}}.  To resolve the pinching issue we proposed to accompany the near EVH near horizon limit by a particular $G_4\rightarrow 0$ limit. Explicitly, we proposed the following triple scaling limit:
horizon area, Hawking temperature and $G_4 \rightarrow0$, keeping the ratios horizon area to temperature and  horizon area to Newton coupling constant fixed.  It implies a particular duality
between 2d CFTÕs on a cylinder and its orbifold: 2d CFT with central charge $c$ on cylinder $R ? S^1$ is dual to 2d CFT with central charge $c K$ on$ R \times S^1/Z_K$ in the large $K$ limit.  It is interesting to elaborate this  issue further.

Let us also comment on the connection between the 2d CFT description we discussed in the previous section and  the Kerr/CFT proposal. A possible connection between these two can come along the lines of \cite{DLCQ-CFT} and discussed with some details in\cite{SheikhJabbaria:2011gc}: The EVH/CFT in the DLCQ description reproduces Kerr/CFT. For the above to work one should, however, extend the validity of our EVH/CFT proposal beyond the strict near EVH region. In other words, generic extremal black holes may be viewed as excitations above the EVH black hole, when one sector of the dual 2d CFT has been excited.

\section*{Acknowledgment}
I would like to thank Eoin \'O Colg\'ain and Shahin Sheikh-Jabbari for useful discussion and comments.  This work was supported by the National Research Foundation of Korea Grant funded by the Korean Government (NRF-2011-0023230).

\end{document}